\pageno=1

\magnification=\magstep1

\def\phi{\emptyset}

\def\zone{2 {\bf Z}}
\def\zp{{\bf Z}_+}
\def\zpo{{\bf Z}}

\def\dur{1}
\def\dk{2}
\def\gri{3}
\def\inui{4}
\def\slt{5}
\def\slf{6}
\rightline{}
\par
\vskip 1.0cm 
\centerline{\bf Survival Probabilities for Discrete Time Models in One Dimension}
\par
\vskip 2.0cm
\centerline{Makoto KATORI}
\par
\vskip 0.1cm
\centerline{{\it Department of Physics, Faculty of Science and Engineering,}}
\centerline{{\it Chuo University, Kasuga, Bunkyo-ku, Tokyo, 112-8551, JAPAN}}
\centerline{{\it katori@phys.chuo-u.ac.jp}}
\par
\vskip 0.3cm
\centerline{Norio KONNO}
\vskip 0.3cm
\par
\centerline{{\it Department of Applied Mathematics, Faculty of Engineering,}}
\centerline{{\it Yokohama National University, Hodogaya-ku, Yokohama, 240-8501, JAPAN}}
\centerline{{\it norio@mathlab.sci.ynu.ac.jp}}
\par
\vskip 0.3cm
\centerline{Hideki TANEMURA}
\par
\vskip 0.3cm
\centerline{{\it Department of Mathematics and Informatics, Faculty of Science,}}
\centerline{{\it Chiba University, Yayoi, Inage-ku, Chiba, 263-8522, JAPAN}}
\centerline{{\it tanemura@math.s.chiba-u.ac.jp}}
\par
\par
\vskip 1.0cm
\noindent  
{\bf Abstract.} We consider survival probabilities for the discrete time process in one dimension, which is known as the Domany-Kinzel model. A convergence theorem for infinite systems can be obtained in the nonattractive case.
\par
\
\par
\
\par
\noindent
{\it Abbreviated title. $\>$} Survival Probabilities for Discrete Time Models.  
\par
\
\par\noindent
All correspondence concerning this paper should be addressed to:
\par\noindent
Dr. Norio KONNO
\par\noindent
Department of Applied Mathematics, Faculty of Engineering, Yokohama National University, 
Tokiwadai, Hodogaya, Yokohama, 240 JAPAN
\par\noindent
Tel and Fax: +81-45-339-4205, e-mail: norio@mathlab.sci.ynu.ac.jp
\par

\vfill\eject
\noindent  
{\bf 1. INTRODUCTION}
\par
\vskip 0.33cm 
In this paper we consider the following one-dimensional discrete-time process $\xi^A _n$ at time $n$ starting from $A \subset {\bf 2Z}$ whose evolution satisfies:
\par\noindent
(i) $P( x \in \xi^A _{n+1} | \xi^A _n ) = f ( |\xi^A _n \cap \{x-1, x+1\}|)$,
\par\noindent
(ii) given $\xi^A _n$, the events $\{ x \in \xi^A _{n+1} \}$ are independent,
where
$$ f(0)=0, \qquad f(1)=p_1, \qquad \hbox{and} \qquad f(2)=p_2,$$ 
with $p_1, \> p_2 \in [0,1].$ This process can be considered on a space $S= \{ s=(x,n)\in \zpo \times \zp : x+n= \hbox{even} \},$ where $\zp = \{0,1,2, \ldots \}.$ See pp.90-98 in Durrett$^{(\dur)}$ for details. This class was first studied by Domany and Kinzel,$^{(\dk)}$ so it is often called the Domany-Kinzel model. 
\par
The oriented bond percolation ($p_1 =p, \> p_2 = 2p - p^2$) and the oriented site percolation 
($p_1 = p_2 = p$) are special cases. The mixed site-bond oriented percolation with the 
probability of open site $\alpha$ and with the probability of open bond $\beta$ corresponds 
to the case of $p_1 = \alpha \beta$ and $p_2 = \alpha ( 2 \beta - \beta^2).$ The reader is referred to ($\dur$).
\par
When $0 \le p_1 \le p_2 \le 1,$ this process is called ``attractive" and has the following nice property: if $\xi^A _n \subset \xi^B_n,$ then we can guarantee that $\xi^A _{n+1} \subset \xi^B _{n+1}$ by using an appropriate coupling.
\par
In the present paper, we study the existence and expression of survival probabilities for this process. In particular, the nonattractive case is interesting, since little results are known on this problem. From now on we will explain the background of our research. Let $Y = \{ A \subset \zone : |A| < \infty \}$ where $|A|$ is the cardinality of $A$. 
\par
The first fundamental fact is that for any $p_1, \> p_2 \in [0,1]$ and $A \subset \zone,$ 
$$ \lim_{n \to \infty}  P ( \xi^A _{2n} \ne \phi \> )$$
exists, since $\phi$ is an absorbing set.  
\par
The second fact is given in the attractive case. That is, for any $0 \le p_1 \le p_2 \le 1$ and $A \subset \zone, \> B \in Y,$ 
$$ \lim_{n \to \infty}  P ( \xi^{A} _{2n} \cap B \ne \phi \> )$$
exists. This result follows from the following ``complete convergence theorem": for any $A \subset \zone,$
$$  \xi^A _{2n}  \Rightarrow P(\tau^A < \infty) \delta_{\phi} + P(\tau^A = \infty) 
\xi^{\zone} _{\infty} \qquad \hbox{as} \qquad  n \to \infty \eqno(1.1)$$
where $\Rightarrow$ means weak convergence, $\tau^A = \inf \{ n : \xi^A _{2n} = \phi \}, \> \delta_{\phi}$ is the pointmass on $\emptyset,$ and a limit $\xi^{\zone} _{\infty}$ is a stationary distribution for the attractive process. 
The above complete convergence theorem can be obtained by similar arguments for the lemma in Griffeath$^{(\gri)}$ (see also Section 5c of Durrett$^{(\dur)}$) which treated a continuous time version. Furthermore, as an immediate consequence of this theorem, we have
$$ \lim_{n \to \infty}  P ( \xi^A _{2n} \cap B \ne \phi \> ) =  P(\tau^A = \infty) P( \xi^{\zone} _{\infty} \cap B \ne \phi \> ) \eqno(1.2)$$
for any $A \subset \zone, \> B \in Y.$ 
In order to clarify the above observation, here we introduce the following notation on survival probabilities starting from $A$ on $B$ by
$$\sigma (A, B)
= \lim_{n \to \infty}  P ( \xi^A _{2n} \cap B \ne \phi \> ) \eqno(1.3) $$
if the right-hand side exists. Using the notation given by (1.3), we can rewrite the above mentioned facts as follows: 
\par\noindent
(i) If $p_1, \> p_2 \in [0,1]$ and $A \subset \zone,$ then $\sigma (A, \zone)$ exists.
\par\noindent
(ii) If $0 \le p_1 \le p_2 \le 1$ (attractive case) and $A \subset \zone, \> B \in Y,$ then $ \sigma (A, B)$ exists, in particular, $\sigma (\zone, B)$ exists. Moreover  
$$ \sigma (A, B) = \sigma (A, \zone) \sigma (\zone, B). \eqno(1.4)$$
\par\noindent
Therefore, it is natural to ask whether or not $\sigma (\zone, A)$ exists if $A$ is finite even in the nonattractive case. The next main theorem gives not only an affirmative answer but also an expression of $\sigma (\zone , A)$ by using $\sigma (D, \zone)$ with $D \subset A$ in a more general setting, i.e. $p_1 \in [0,1), \>p_2 \in (0,1]$ and $p_2 < 2 p_1.$  
\par
\
\par
\proclaim Theorem 1. We assume that $p_1 \in [0,1), \> p_2 \in (0,1]$ and $p_2 < 2 p_1.$ For any $A \in Y,$  
$$
\sigma (\zone , A) = \sum_{D \subset A, D \not= \phi} \alpha^{|D|} (1- \alpha)^{|A 
\setminus D|} \sigma (D, \zone), \eqno(1.5)
$$
where $\alpha = p_1 ^2 /(2p_1 - p_2).$ In particular, if $A = \{x \},$ then
$$
\sigma (\zone , \{ x \}) = \alpha \sigma (\{ x \} , \zone).
$$
If $A = \{x,y \}$ with $x \not= y,$ then
$$
\sigma (\zone , \{ x,y \}) = 2 \alpha (1- \alpha) \sigma (\{ x \} , \zone)
+ \alpha^2 \sigma (\{ x , y \} , \zone).
$$ 
\par
\
\par
We should remark that it is easy to see that the process with $p_1 \in [0,0.5]$ and $p_2 \in [0,1]$ starting from a finite set dies out by comparison with a branching process $Z_n$ as follows. Each particle gives rise to $Y$ particles in the next generation where $Y$ is given by
$ P( Y=2 )=p_1 ^2, \> P( Y=1 )=2p_1(1-p_1), \> P( Y=0 )=(1-p_1)^2.$ For details, see p.97 in Durrett$^{(\dur)}$.
\par
In the case of $p_2 \ge 2 p_1$, the process is attractive ($p_2 \ge p_1$), and $\sigma (\zone , A) = 0$ if $A$ is finite (see the above discussion), so this case is not interesting. On the other hand, the above theorem implies that for any finite $A,$ $\sigma (\zone , A) $ exists in the case of $p_2 < 2 p_1$ except for $p_1=1$ or $p_2=0$. The crucial point is that this result treats even the nonattractive case ($p_2 < p_1$). 
\par
Conversely, from Theorem 1, we can express $\sigma (A, \zone)$ by using $\sigma (\zone, D)$ with $D \subset A.$ To see this, we put 
$$
\eqalign{
\rho (\zone, A) &= (1/(1-\alpha))^{|A|} \sigma (\zone ,A), \cr
\rho (A, \zone) &= (\alpha/(1-\alpha))^{|A|} \sigma (A,\zone). 
\cr}$$
Then from (1.5) we have
$$ \rho (\zone,A) = \sum_{D \subset A, D \not= \phi} \rho (D, \zone),$$
and so by the M\"obius transformation we can obtain
$$ \rho (A, \zone) = \sum_{D \subset A, D \not= \phi} (-1)^{|A|-|D|} \rho (\zone,D).$$
Hence
$$
\eqalign{
\sigma (A, \zone) 
&= ((1-\alpha)/\alpha)^{|A|}
\sum_{D \subset A, D \not= \phi} (-1)^{|A|-|D|} (1/(1-\alpha))^{|D|} \sigma (\zone, D)
\cr
&=
\sum_{D \subset A, D \not= \phi} (1/\alpha)^{|D|} ((\alpha -1)/ \alpha)^
{|A|-|D|} \sigma (\zone, D)
\cr
}$$
This gives the desired expression: for any $A \in Y,$
$$
\sigma (A, \zone) = \sum_{D \subset A, D \not= \phi} (1/\alpha)^{|D|} ((\alpha -1)/ \alpha)^
{|A \setminus D|} \sigma (\zone, D), \eqno(1.6)
$$
\par
In the proof of Theorem 1, we introduce a new construction of the process using a signed measure with $\alpha = p_1^2 /(2p_1 - p_2)$ and $\beta = 2 - p_2/p_1$. More detailed discussions will be found in the next section which is devoted to the proof of Theorem 1.
\par
>From this theorem we can immediately get the following convergence theorem which applies to some nonattractive models. (The standard argument can be found in page 71 of ($\dur$).)
\par
\
\par 
\proclaim Corollary 2. We assume that $p_1 \in [0,1), \> p_2 \in (0,1]$ and $p_2 < 2 p_1.$ 
Then we have
$$  P(\xi^{\zone} _{2n} \in \cdot ) \Rightarrow \mu \qquad \hbox{as} \qquad  n \to \infty, $$
where $\mu$ is the translation invariant probability measure such that
$$  \mu ( \xi \cap A \not= \phi) = \sum_{D \subset A, D \not= \phi} \alpha^{|D|} (1- \alpha)^{|A 
\setminus D|} \sigma (D, \zone), $$
for any $A \subset \zone$ with $|A| < \infty.$ 
\par
\
\par\noindent
We should remark that $\zone$ in $\xi^{\zone} _{2n}$ in Corollary 2 can be replaced by $B \subset \zone$ with $|\zone \setminus B| < \infty,$ and a random set $B$ which is Bernoulli distributed with parameter $\theta \in (0,1],$ by modifying our proofs slightly. 
\par
Moreover, combining (1.4) with $A=\{x\}, \> B=\{y\}$ and Theorem 1, we get 
the following relation between local survival probability $\sigma (\{ x \}, \{y \})$ and global survival probability $\sigma (\{x\}, \zone).$ Note that $\sigma (\{x\}, \zone)=\sigma (\{y\}, \zone)$ for any $x,y \in \zone.$ 
\par
\
\par
\proclaim Corollary 3. We assume that $0 \le p_1 \le p_2 < 2 p_1.$ Let $\alpha = p_1 ^2 /(2p_1 - p_2).$ For any $x,y \in \zone,$
$$ \sigma (\{ x \}, \{y \}) = \alpha \sigma (\{x\}, \zone)^2, \eqno(1.7)$$
In particular, when $p_2 = 2 p_1 - p_1 ^2$ (oriented bond percolation), then we have
$$ \sigma (\{ x \}, \{y \}) = \sigma (\{x\}, \zone)^2, \eqno(1.8)$$
and, when $p_2 =  p_1$ (oriented site percolation), then we have
$$ \sigma (\{ x \}, \{y \}) = p_1 \sigma (\{x\}, \zone)^2. \eqno(1.9)$$
\par
\
\par\noindent
>From (1.7), we see that $\sigma (\{ x \}, \{y \})$ is independent 
of $y$. Recently Inui {\it et al.} $^{(\inui)}$ conjectured special cases of the above 
results, i.e. (1.8) and (1.9) with $x=y$, by Pad\'e approximants of series expansion. Therefore, to prove (1.9) was one of our motivations of the present paper. Concerning (1.8), this equality can be easily obtained by the complete convergence theorem as in the case of the basic contact process, since oriented bond percolation is self-dual.
\par
Theorem 1 can be regarded as a generalization of time-reversal duality in limit of time $n \to \infty .$ It should be noted that this self-duality holds for any $p_1 \in [0,1), \> p_2 \in (0,1],$ and $p_2 < 2 p_1$ in our theorem. Another generalization of duality for continuous-time and nonattractive models has been studied by Sudbury and Lloyd$^{(\slt , \slf)}$. In their results, however, self-duality holds only for special cases.
\par

\vskip 0.5cm
\par\noindent  
{\bf 2. PROOF OF THEOREM 1}
\vskip 0.33cm
\par
First we introduce these spaces:
$$
\eqalign{
S &= \{ s=(x,n)\in \zpo \times \zp : x+n= \hbox{even} \},
\cr
B &= \{ b=((x,n), (x+1,n+1)),((x,n), (x-1,n+1)) : (x,n)\in S \}
\cr
X(S) &= \{0,1\}^S, \quad X(B)= \{0,1\}^B, \quad
X=X(S)\times X(B),
\cr }
$$
where $\zp = \{0,1,2, \ldots \}.$ For given $\zeta = (\zeta_1, \zeta_2) \in X$,
we say that $s = (y, n+k)\in S$ can be reached from 
$s' =(x, n)\in S$ and write $s'\to s$, 
if there exists a sequence $s_0,s_1,\dots, s_k$ of members of $S$
such that
$s'=s_0$, $s=s_k$ and
$\zeta_1(s_i)=1, i=0,1,\dots,k$, 
$\zeta_2((s_i,s_{i+1}))=1, i=0,1,\dots,k-1$.
We also say that  
$G\subset S$ can be reached from $G'\subset S$ 
and write $G'\to G$,
if there exist $s\in G$ and $s'\in G'$ such that 
$s'\to s$.

We introduce the signed measure $m$ on $X$ defined by
$$
m(\Lambda) = \alpha^{k_1} (1-\alpha)^{j_1} \beta^{k_2} (1-\beta)^{j_2},
$$
for any cylinder set 
$$
\eqalign{
\Lambda 
=\{ (\zeta_1, \zeta_2) \in X : 
& \> \zeta_1(s_i)=1,i=1,2,\dots ,k_1,
\> \zeta_1(s'_i)=0,i=1,2,\dots ,j_1,
\cr
&\> \zeta_2(b_i)=1,i=1,2,\dots ,k_2,
\> \zeta_2(b'_i)=0,i=1,2,\dots ,j_2 \},
\cr}
$$
where $s_1,\dots,s_{k_1}, s'_1,\dots,s'_{j_1}$ 
are distinct elements of $S$ and $b_1,\dots,b_{k_2}, b'_1,\dots,b'_{j_2}$ 
are distinct elements of $B$,
and 
$\alpha = p_1^2 /(2p_1 - p_2)$,
$\beta = 2 - p_2/p_1$. 
\par
If $p_2 < 2 p_1$ and $p_2 > 2p_1 - p_1 ^2,$ then $\alpha > 1$ and $\beta \in (0,1).$ If $p_2 \le 2p_1 - p_1 ^2$ and $p_2 \ge p_1,$ then $\alpha, \beta \in [0,1].$ This case corresponds to the mixed site-bond oriented percolation with the probability of open site $\alpha$ and with the probability of open bond $\beta$ where $p_1 = \alpha \beta$ and $p_2 = \alpha ( 2 \beta - \beta^2).$ That is why we choose $\alpha = p_1^2 /(2p_1 - p_2)$ and $\beta = 2 - p_2/p_1$ in our construction. Moreover, if $p_2 < p_1,$ then $\alpha \in (0,1)$ and $\beta \in (1,2].$ From the above observation, we see that the measure becomes signed measure in the first and third cases, since $\alpha >1$ and $1- \beta <0$ respectively.
Concerning the relation between the original models with $p_1$ and $p_2$ and this construction with $\alpha$ and $\beta$, see Figure 1. 
\par
We define the conditional signed measures as follows: 
$$
\eqalign{
m_k(\cdot)&= m(\cdot | \zeta_1(s)=1, s \in S_k ),
\cr
m_{k,j}(\cdot)&= m(\cdot | \zeta_1(s)=1, s \in S_k \cup S_j ),
\cr}
$$
where $S_k = \{ (x,n) \in S : n=k \}$.
Then by simple observation we see that
$$
P(\xi_n^A \ni y) = m_0( \{0\}\times A \to (y,n)).
$$
For a fixed even nonnegative number $k$,
we introduce the map $r_k$ from $S$ to $S$ defined by
$$
r_k(x,n)= 
\cases{ 
(x,k-n), &$n=0,1,\dots,k$ ;\cr
 (x,n),  &otherwise, \cr}
$$
and the map $R_k$ from $X$ to $X$ defined by
$$
R_k \zeta = ((R_k \zeta)_1, (R_k \zeta)_2));
(R_k \zeta)_1(s) = \zeta_1(r_k s),
(R_k \zeta)_2((s,s')) = \zeta_2((r_k s', r_k s)).
$$
Note that $m$ is $R_k$-invariant.
To prove Theorem 1 we use the following lemma. This is a trivial Markov chain fact, however for the convenience of the reader, we present the proof.

\vskip 1.0cm

\par
\proclaim Lemma 4.
Suppose that $p_1 \in [0,1)$ and $p_2 \in [0,1]$.
Then for any positive integer $\ell$ 
and $A \subset \zone$,
we have
$$
\lim_{n\to\infty} 
P( 1 \le |\xi_n^A | \le \ell, \Omega_{\infty}^A )
=0
$$
where $\Omega_{\infty} ^A = \{\xi_n^A \not= \emptyset
 \> \hbox{for any} \> n \ge 0 \}.$ 
\par
\noindent
Proof.
We can choose $\varepsilon = \varepsilon (p_1, p_2, \ell) >0$
such that 
$$
P( 1 \le |\xi_n^A | \le \ell, \xi_{n+\ell}^A = \emptyset)
\ge \varepsilon 
P( 1 \le |\xi_n^A | \le \ell)
$$
Then
$$
\eqalign{
P( 1 \le |\xi_n^A | \le \ell, \Omega_{\infty}^A )
&\le 
P( 1 \le |\xi_n^A | \le \ell, \xi_{n+\ell}^A \not= \emptyset)
\cr
&\le
(1-\varepsilon)
P( 1 \le |\xi_n^A | \le \ell)
\cr}
$$
Hence
$$
\eqalign{
\varepsilon 
P( 1 \le |\xi_n^A | \le \ell, \Omega_{\infty}^A )
&\le (1-\varepsilon)
P( 1 \le |\xi_n^A | \le \ell, ({\Omega_{\infty}^A})^c )
\cr
&\le (1-\varepsilon)
P( \xi_n^A \not= \emptyset, ({\Omega_{\infty}^A})^c ).
\cr}
$$
Therefore
$$
\lim_{n\to\infty} 
P(1 \le |\xi_n^A | \le \ell, \Omega_{\infty}^A )
\le \lim_{n\to\infty} 
{{1-\varepsilon}\over {\epsilon}}
P( \xi_n^A \not= \emptyset, ({\Omega_{\infty}^A})^c )=0.
$$
This completes the proof.

\vskip 1.0cm

Now we prove Theorem 1. Suppose that $n$ is even. To begin, we observe
$$
\eqalign{
&P( \xi_n^{\zone} \cap A \not= \phi ) 
\cr
&= m_0(S_0 \to A \times \{ n\})
\cr
&=\sum_{D \subset A, D \not= \emptyset}
m_{0}(S_0 \to A \times \{n\}, \> \zeta_1 ((x,n)) =1 \> \hbox{for any} \> x \in D, \cr
& \qquad \qquad \qquad \qquad \qquad 
\zeta_1 ((x,n)) =0 \> \hbox{for any} \> x \in A \setminus D)
\cr
&=\sum_{D \subset A, D \not= \emptyset}
m_{0}(S_0 \to D \times \{n\}, \> \zeta_1 ((x,n))=1 \> \hbox{for any} \> x \in D, \cr
& \qquad \qquad \qquad \qquad \qquad 
\zeta_1 ((x,n)) =0 \> \hbox{for any} \> x \in A \setminus D)
\cr
&=\sum_{D \subset A, D \not= \emptyset}
m_{0,n}(S_0 \to D \times \{n\}) \alpha^{|D|} (1-\alpha)^{|A \setminus D|}.
\cr
}$$
Since $m$ is $R_n$-invariant, $m_{0n}$ is also $R_n$-invariant. So we see that
$$
\eqalign{
& m_{0,n}(S_0 \to D \times \{n\}) \cr
&= m_{0,n}(D \times \{0\} \to S_n) \cr
&= 
\sum_{C: 0<|C|< \infty}
m_{0,n}(D \times \{0\} \to (x,n-1) \> \hbox{for any} \> x \in C, \cr
& \qquad \qquad \qquad \qquad 
D \times \{0\} \not\to C^c \times \{n-1\}, \> C \times \{n-1\} \to S_n)
\cr
&=
\sum_{C: 0<|C|< \infty}
m_{0}(\{0\}\times D \to (x,n-1) \> \hbox{for any} \> x \in C, \cr
& \qquad \qquad \qquad \qquad 
D \times \{0\} \not\to C^c \times \{n-1\}) \times 
m_{n-1,n}(C \times \{n-1\} \to S_n)
\cr
&=
\sum_{C: 0<|C|< \infty} P( \xi^D _{n-1} = C) \left\{ 1 - (1 - \beta)^{2|C|} \right\}
\cr
&=
P( \xi^D _{n-1} \not= \phi) -  
E[(1-\beta)^{2 |\xi^D _{n-1}| }; \xi^D _{n-1} \not= \emptyset].
\cr
}$$
where $G' \not\to G \> (G, G' \subset S)$ means that there exist no $s\in G$ and $s'\in G'$ such that $s'\to s$. Hence, to obtain (1.5) it is enough to show
$$
\lim_{n \to \infty}
E[(1-\beta)^{2 |\xi^D _{n-1} |}; \xi^D _{n-1} \not= \emptyset] =0. 
\eqno(2.1)
$$
For any $\ell \in \{1,2, \ldots \},$ we see that
$$
\eqalign{
& E[(1-\beta)^{ 2 |\xi^D _{n-1} | }; \xi^{D} _{n-1} \not= \emptyset] \cr
&\le E[(1-\beta)^{ 2 |\xi^{D} _{n-1} |}; \Omega^D _{\infty}]
+ 
P(\xi^{D} _{n-1} \not= \emptyset, (\Omega_{\infty}^{D})^c)
\cr
&\le
P(1 \le |\xi_{n-1}^{D}| \le \ell, \Omega_{\infty}^{D} )
+ 
(1-\beta)^{2 \ell}
+
P(\xi_{n-1}^{D} \not= \emptyset, ( \Omega_{\infty}^{D})^c )
\cr
}\eqno(2.2)$$
Note that $\{ \xi^{D} _{n-1} \not= \emptyset \} \setminus \Omega^{D} _{\infty}$
converges to the empty set as $n \to \infty$. So, by (2.2) and Lemma 4 we have
$$
\lim_{n \to \infty}
E \left[ (1-\beta)^{2 | \xi^D _{n-1}| }; \xi^D _{n-1} \not= \emptyset \right] 
\le 
(1-\beta)^{2 \ell}, 
$$
for any $\ell \in \{1,2, \ldots \}$. Under the condition that $0 < p_2 < 2p_1,$ we have 
$1 - \beta = p_2 /p_1 - 1 \in (-1,1).$ Therefore, letting $\ell \to \infty$ gives the 
desired result (2.1).
\par

\vskip 1.0cm
\noindent
{\bf References}
\par
\
\par
\item{$\dur$.} R. Durrett, {\it Lecture Notes on Particle Systems and Percolation\/} 
(Wadsworth, Inc., California, 1988).

\item{$\dk$.} E. Domany and W. Kinzel, Equivalence of cellular automata to Ising models and directed percolation. {\it Phys. Rev. Lett.$\>$} {\bf 53}:311-314 (1984).

\item{$\gri$.} D. Griffeath, Limit theorems for nonergodic set-valued Markov processes. 
{\it Ann.Probab.$\>$} {\bf 6}:379-387 (1978).

\item{$\inui$.} N. Inui, N. Konno, G. Komatsu and K. Kameoka, Local directed percolation probability in two dimensions. {\it J. Phys. Soc. Jpn.$\>$} {\bf 67}:99-102 (1998).

\item{$\slt$.} A. Sudbury and P. Lloyd, Quantum operators in classical probability theory: II. The concept of duality in interacting particle systems. {\it Ann. Probab.$\>$} {\bf 23}:1816-1830 (1995).

\item{$\slf$.} A. Sudbury and P. Lloyd, Quantum operators in classical probability theory: IV. Quasi-duality and thinnings of interacting particle systems. {\it Ann. Probab.$\>$} {\bf 25}:96-114 (1997).

\vfill\eject
\noindent
{\bf Figure Captions}
\par
\
\par\noindent
Figure 1. The relation between the original models with
$p_{1}$ and $p_{2}$ and the construction with
$\alpha$ and $\beta$.
The open (resp. closed) sites with $\zeta_{1}(s)=1$ (resp. 0)
are denoted by full (resp. open) circles and the
open (resp. closed) bonds with $\zeta_{2}(b)=1$ (resp. 0)
are denoted by full (resp. broken) lines. The configurations
on the enclosed sites are given.
We consider all the cases in which the site below is open
and it is connected to one of the enclosed open sites
by at least one open bond. For each open site
put $\alpha$ and for each open (resp. closed) bond put
$\beta$ (resp. $1-\beta$).

\par
\end